\documentclass[aps,twocolumn,a4paper,floatfix,superscriptaddress,prapplied,longbibliography]{revtex4-2}
\usepackage[toc,page]{appendix}
\usepackage{braket}
\usepackage{epsfig,graphicx,times}
\usepackage{amstext}
\usepackage{amsmath}
\usepackage{amssymb}
\usepackage{mathrsfs}
\usepackage{dcolumn}
\usepackage{bm}
\usepackage[tight]{subfigure}
\usepackage[colorlinks,linkcolor=blue,anchorcolor=blue,citecolor=blue,urlcolor=blue]{hyperref}
\usepackage{color}
\usepackage{siunitx}
\usepackage{appendix}
\usepackage{placeins}
\usepackage{textcomp}
\usepackage{bbold}
\usepackage{float}
\usepackage{verbatim}
\usepackage{minitoc}
\usepackage[dvipsnames]{xcolor}

\usepackage{soul}
\usepackage[framemethod=tikz]{mdframed}

\begin{document}

\title{Phase-controlled asymmetric optomechanical entanglement against optical backscattering}
\author{Jing-Xue Liu}
\affiliation{Key Laboratory of Low-Dimensional Quantum Structures and Quantum Control of Ministry of Education, \\
Department of Physics and Synergetic Innovation Center for Quantum Effects and Applications, \\
Hunan Normal University, Changsha 410081, China}

\author{Ya-Feng Jiao}\email{yfjiao@hunnu.edu.cn}
\affiliation{Key Laboratory of Low-Dimensional Quantum Structures and Quantum Control of Ministry of Education, \\
Department of Physics and Synergetic Innovation Center for Quantum Effects and Applications, \\
Hunan Normal University, Changsha 410081, China}
\affiliation{Laboratory of Chemical Biology $\&$ Traditional Chinese Medicine Research, \\
Ministry of Education College of Chemistry and Chemical Engineering, \\
Hunan Normal University, Changsha 410081, China}

\author{Ying Li}
\affiliation{Key Laboratory of Low-Dimensional Quantum Structures and Quantum Control of Ministry of Education, \\
Department of Physics and Synergetic Innovation Center for Quantum Effects and Applications, \\
Hunan Normal University, Changsha 410081, China}

\author{Xun-Wei Xu}
\affiliation{Key Laboratory of Low-Dimensional Quantum Structures and Quantum Control of Ministry of Education, \\
Department of Physics and Synergetic Innovation Center for Quantum Effects and Applications, \\
Hunan Normal University, Changsha 410081, China}

\author{Qiong-Yi He}
\affiliation{State Key Laboratory for Mesoscopic Physics, School of Physics, \\Frontiers Science Center for Nano-optoelectronics, $\&$ Collaborative Innovation Center of Quantum Matter, \\Peking University, Beijing 100871, China}
\affiliation{Collaborative Innovation Center of Extreme Optics, Shanxi University, Taiyuan, Shanxi 030006, China}

\author{Hui Jing}\email{jinghui73@gmail.com}
\affiliation{Key Laboratory of Low-Dimensional Quantum Structures and Quantum Control of Ministry of Education, \\
Department of Physics and Synergetic Innovation Center for Quantum Effects and Applications, \\
Hunan Normal University, Changsha 410081, China}
\affiliation{Synergetic Innovation Academy for Quantum Science and Technology, Zhengzhou University of Light Industry, Zhengzhou 450002, China}

\date{\today}

\begin{abstract}	
Quantum entanglement plays a key role in both understanding the fundamental aspects of quantum physics and realizing various quantum devices for practical applications. Here we propose how to achieve coherent switch of optomechanical entanglement in an optical whispering-gallery-mode resonator, by tuning the phase difference of the driving lasers. We find that the optomechanical entanglement and the associated two-mode quantum squeezing can be well tuned in a highly asymmetric way, providing an efficient way to protect and enhance quantum  entanglement against optical backscattering, in comparison with conventional symmetric devices. Our findings shed a new light on improving the performance of various quantum devices in practical noisy environment, which is crucial in such a wide range of applications as noise-tolerant quantum processing and the backscattering-immune quantum metrology.
\end{abstract}

\maketitle

\section{Introduction}\label{Int}
Cavity optomechanics (COM)~\cite{Kippenberg2007,Verhagen2012,aspelmeyer2014Cavityb,Xiong2015,Barzanjeh2022}, based on radiation-pressure mediated coherent light-motion coupling, have become a cornerstone in many important quantum applications, such as microwave-to-light quantum conversion~\cite{Forsch2020,Sahu2022,Hoenl2022}, COM-based quantum sensing~\cite{McClelland2011,Massel2012,Zhao2020,Qvarfort2018}, and phonon lasing~\cite{Cui2021}, to name only a few. An emerging field, which is important for both fundamental studies of quantum physics and applications in quantum information science, is COM-based creation and engineering of quantum entanglement~\cite{Sarma2021a,vitali2007Optomechanical,Genes2008,Ghobadi2014,Liu2021aa,Li2018abcd,Karg2020}. By using microwave circuits or COM crystals, entangled states have been achieved for electromagnetic fields and mechanical motion~\cite{palomaki2013Entangling,Riedinger2016,Marinkovic2018}. Very recently, macroscopic quantum correlations were demonstrated even between a laser and a $40\,\textrm{kg}$ mirror at room temperature~\cite{Yu2020}, which can be used for improving the performance of gravitational wave detectors. Also, COM devices have been used to create quantum entanglement between propagating optical fields~\cite{barzanjeh2019Stationarya,chen2020Entanglementa} or between massive mechanical elements~\cite{Mancini2002,Huang2009,Tan2013,Li2015,OckeloenKorppi2018,riedinger2018Remote,mercierdelepinay2021Quantum,kotler2021Direct}. For example, high-fidelity entanglement of distinct mechanical motions was deterministically generated in a multi-tone-laser driven microwave COM system~\cite{mercierdelepinay2021Quantum,kotler2021Direct}. However, for practical uses in a noisy environment or with material defects, a long-standing challenge is how to protect and efficiently engineer the fragile quantum states of COM devices.

Another field which is closely related to the present work is synthetic gauge fields, proposed and achieved in neutral systems such as atomic gases~\cite{Cui2013ab,Li2022,Goldman2014}, photonic devices~\cite{Yang2019,Ozawa2019,Fang2012a,Chen2012ab}, and acoustic or COM systems~\cite{Yang2021a,Yang2016a,Yang2017ac,Longhi2015}. These synthetic fields can be useful in e.g., simulating many-body physics or achieving topological control~\cite{Huang2022,Mittal2014,Ozawa2019d,Chalabi2019,Hauke2012}. In a very recent experiment~\cite{chen2021Synthetica}, by tuning a phase-controlled COM resonator, synthetic gauge field was demonstrated and used to achieve a nonreciprocal optical transmission. Inspired by this experiment~\cite{chen2021Synthetica}, here we study how to engineer purely quantum COM effects, instead of classical transmission rates, by using such a phase-controlled COM system. We note that this system has also been proposed to achieve optical nonreciprocity~\cite{Yan2019ab,Xu2015} and photon blockade~\cite{Xu2020,Li2022a}. In particular, in Ref.\,\cite{Sun2017}, phase-controlled quantum steering has been explored with blue-detuned driving lasers (relative to the cavity frequency).

In this work, we propose to achieve asymmetric control of steady light-motion entanglement in the red-detuned regime, by using such a phase-controlled COM system~\cite{chen2021Synthetica}, and then we show its robustness against optical backscattering. Specifically, we consider an optical whispering-gallery-mode (WGM) resonator supporting two counter-propagating optical modes and a mechanical breathing mode. By driving this system from left and right directions and tuning the phase difference of the driving lasers, we show that COM entanglement can be manipulated in an asymmetric way. In particular, we find that the ability of tuning asymmetry of COM entanglement provides a new way to directionally enhance the entanglement even in the presence of optical backscattering. We note that material imperfections or defects always exist in solid COM devices, leading to optical backscattering and thus usually being harmful for quantum devices. Our work sheds a new light on achieving robust quantum COM devices, which is well within current experimental capabilities and well compatible with other existing techniques, e.g., topological structures~\cite{Rechtsman2016,Wang2019,Gneiting2019,Mittal2016,Wang2009}, weak measurement~\cite{Kim2012,Foletto2020,Man2012,Zhang2015,Lim2014}, polarization modulation~\cite{Jaffe2022}, and nonreciprocal or dark-mode control~\cite{Jiao2020,Lai2022,Kim2019,Tang2022a,Otterstrom2020}.

\begin{figure}
	\centerline{\includegraphics[width=0.48\textwidth]{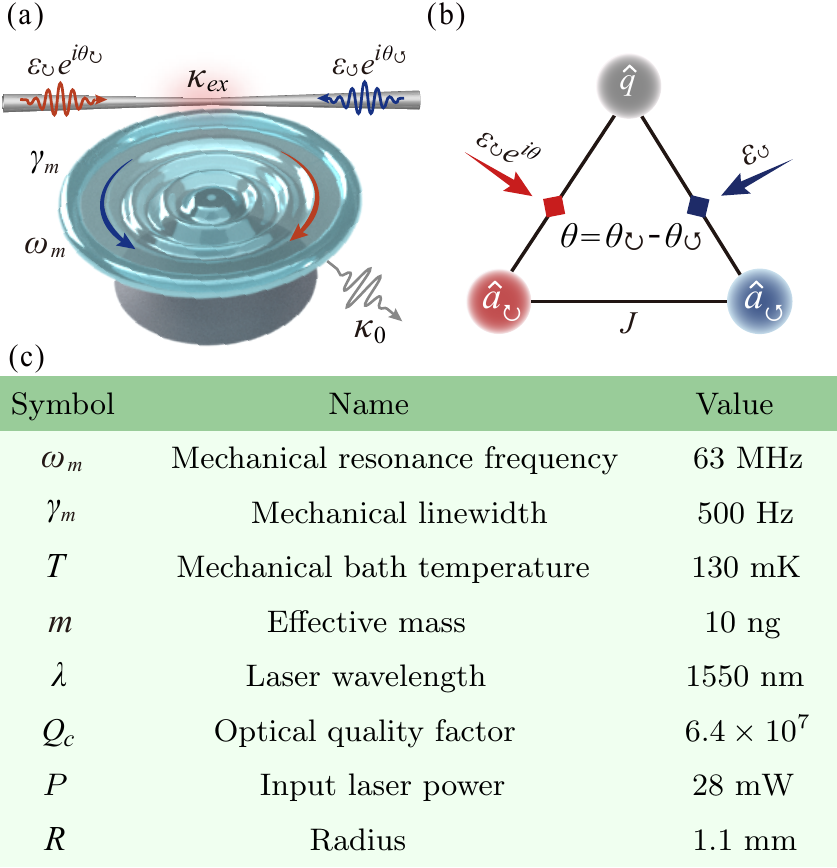}}
	\caption{(a) Schematic diagram of a microdisk WGM resonator with double-pump lasers input from two opposite directions. The WGM resonator supports two degenerate counter-propagating optical modes and a mechanical breathing mode. The driving fields could be coupled into or out of the WGM resonator via an evanescent fiber-cavity coupling. (b) Energy diagram for the mode couplings between CW, CCW and mechanical modes. When light travels for an enclosed loop, a nontrivial phase could be accumulated, thus resulting in a gauge field. (c) Experimentally feasible parameters used in our numerical calculation, which are partially chosen from Refs.~\cite{maayani2018Flying,Mao2022,righini2011Whispering,chen2021Synthetica}.}\label{Fig1}
\end{figure}

\begin{figure*}
	\centerline{\includegraphics[width=0.96\textwidth]{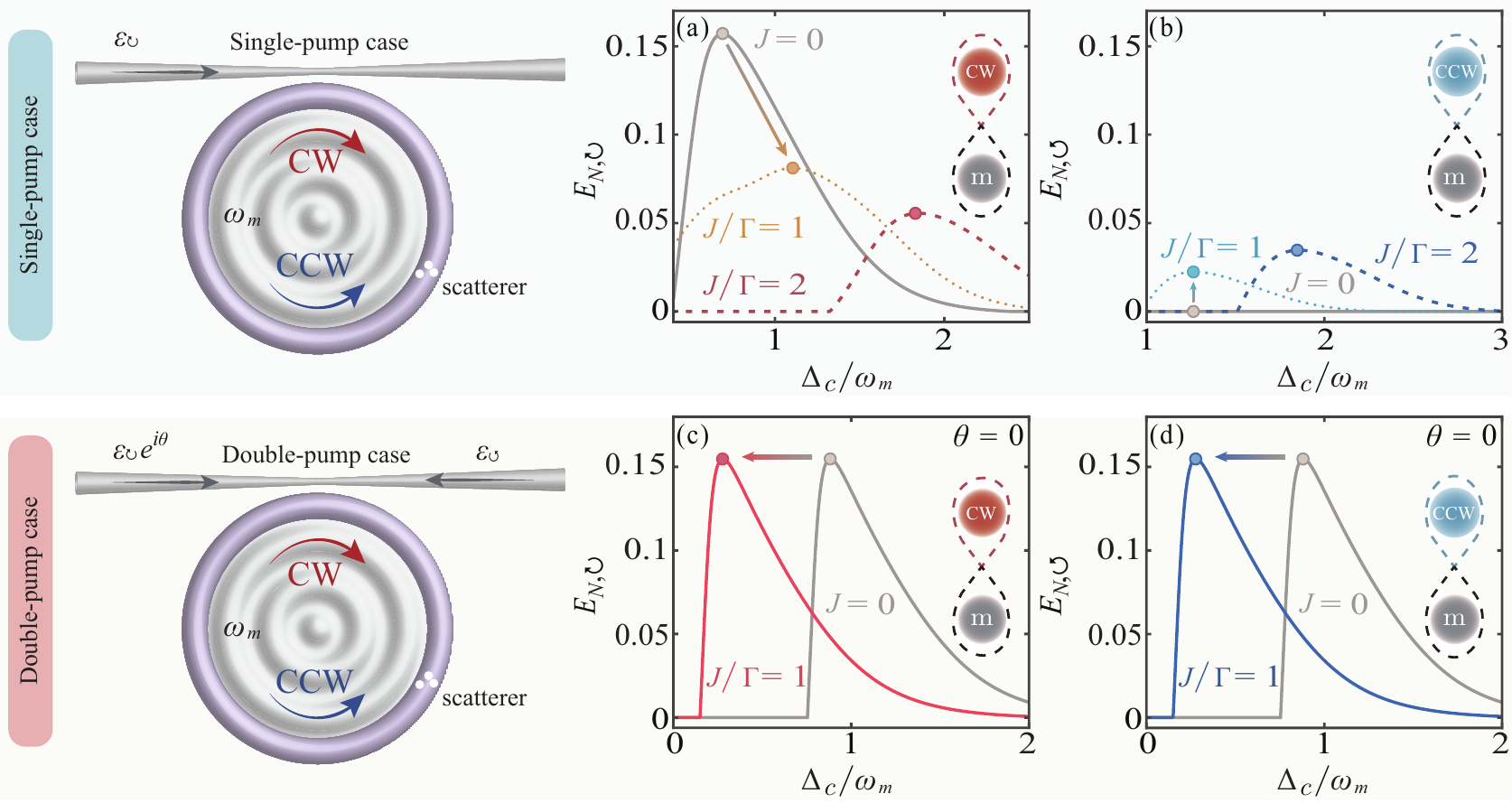}}
	\caption{\label{Fig2}Dependence of COM entanglement on optical backscattering induced couplings with regards to single- or double-pump case. (a)-(b) The logarithmic negativity $E_{N,j}$ versus the scaled optical detuning $\Delta_{c}/\omega_{m}$ for different values of optical coupling rate $J$, with $\varepsilon_{\circlearrowright}\neq0$ and $\varepsilon_{\circlearrowleft}=0$. (c)-(d) The logarithmic negativity $E_{N,j}$ versus the scaled optical detuning $\Delta_{c}/\omega_{m}$ for different values of optical coupling rate $J$, with $\varepsilon_{\circlearrowright}=\varepsilon_{\circlearrowleft}\neq0$, and $\theta=0$. The parameters are chosen as the same in the table of Fig.\,\ref{Fig1}(c).}\label{Fig2}
\end{figure*}

\section{Phase-controlled COM: quantum dynamics}\label{sec:model}

As discussed in detail in Refs.\,~\cite{Berry1984,chen2021Synthetica}, when particles with charge $q$ traveled from $r_{i}$ to $r_{j}$ in a gauge potential $\mathcal{S}$ for an enclosed loop, a nontrivial Peierls phase could be accumulated by path integral, i.e., $\theta_{ij}=(q/\hbar)\int_{r_{i}}^{r_{j}}dr\cdot\mathcal{S}$, which thus results in the gauge fields. For neutral systems, synthetic gauge fields have recently been exprimentally simulated based on real-space lattices in various systems~\cite{Price2022,Dalibard2011,Rechtsman2013}. In a very recent experiment, a controllable synthetic gauge field was also obtained with virtual lattices of bosonic modes in a single COM resonator, and, simultaneously, by regulating the magnetic fluxes of the synthetic gauge field, nonreciprocal conversion between optical and mechanical modes has been observed. In particular, they show that the Peierls phase for their optical systems could be manipulated at will through adjusting the phase difference of the driving lasers (see Ref.\,~\cite{chen2021Synthetica} for more details), which provides a versatile and controllable method to engineer the synthetic gauge field and light-motion interactions.

Inspired by this experiment~\cite{chen2021Synthetica}, here we study how to achieve coherent asymmetric control of COM entanglement by tuning the relative phase of driving lasers, and unveil its robustness against optical backscattering. As shown in Figs.\,\ref{Fig1}(a) and \ref{Fig1}(b), we consider a high-quality WGM microdisk resonator, which supports two counter-propagating optical modes and a mechanical breathing mode. The counter-propagating clockwise (CW) and counterclockwise (CCW) optical modes are degenerate in the resonator, as described by $\omega_{c}$. By evanescently coupling the resonator to an optical fiber and applying two pump lasers from both two sides, the CW and CCW optical modes can be excited and then coupled to the mechanical breathing mode due to the COM interaction. Besides, the imperfections of WGM resonators, such as surface roughness or material inhomogeneous, can cause optical backscattering and thus lead to couplings of CW and CCW modes~\cite{Kippenberg2002}. Therefore, in a rotating frame with respect to drive frequency $\omega_{l}$,  the effective Hamiltonian of this system can be written in the simplest level as
\begin{align}
	\nonumber
	\hat{H}\!=&~\hbar\Delta_{c}\hat{a}_{\scriptstyle{\circlearrowright}}^{\dagger}\hat{a}_{\circlearrowright}\!+\hbar\Delta_{c}\hat{a}_{\circlearrowleft}^{\dagger}\hat{a}_{\circlearrowleft}+\!\dfrac{\hbar\omega_{m}}{2}(\hat{p}^{2}\!+\!\hat{q}^{2}\!)\\ \nonumber
	&+\hbar{J}(\hat{a}_{\circlearrowright}^{\dagger}\hat{a}_{\circlearrowleft}+\hat{a}_{\circlearrowleft}^{\dagger}\hat{a}_{\circlearrowright})\\
	\nonumber
	&-\hbar{G_{0}}(\hat{a}_{\circlearrowright}^{\dagger}\hat{a}_{\circlearrowright}+\hat{a}_{\circlearrowleft}^{\dagger}\hat{a}_{\circlearrowleft})\hat{q}\\
	&+i\hbar(\varepsilon_{\circlearrowright}\hat{a}_{\circlearrowright}^{\dagger}e^{-i\theta_{\circlearrowright}}+\varepsilon_{\circlearrowleft}\hat{a}_{\circlearrowleft}^{\dagger}e^{-i\theta_{\circlearrowleft}}-\mathrm{H.c.}),\label{eq1}
\end{align}
where $\hat{a}_{j}$ $(\hat{a}_{j}^{\dagger})$ is the annihilation (creation) operator for optical modes with $j=\circlearrowright,\circlearrowleft$ indexing CW or CCW direction, $\Delta_{c}=\omega_{c}-\omega_{l}$, and $\hat{q}$ $(\hat{p})$ is the dimensionless mechanical displacement (momentum) operator for a mechanical breathing mode with fundamental frequency $\omega_{m}$. $\textit{J}$ denotes the strength of backscattering induced coupling between CW and CCW modes. Also, $G_{0}=(\omega_{c}/R)\sqrt{\hbar/m\omega_{m}}$ denotes the single-photon COM coupling rate, with $m$ and $R$ the effective mass and radius of the resonator. The phase and amplitude of the driving field are described by $\theta_{j}$ and $|\varepsilon_{j}|=\sqrt{2\kappa_{ex} P_{j}/\hbar\omega_{l}}$, where $P_{j}$ is the input laser power, and $\kappa_{ex}$ is the fiber-cavity coupling rate. The phase difference of the double pump lasers is defined by $\theta\equiv\theta _{\circlearrowright}-\theta_{\circlearrowleft}$. By adjusting this phase difference, a controllable magnetic flux of the synthetic gauge field can be achieved, as already demonstrated in the very recent experiment~\cite{chen2021Synthetica}.

By employing the quantum Langevin equations (QLEs), the equations of motion of each optical and mechanical operators are obtained as
\begin{align}
	\nonumber \dot{\hat{a}}_{\circlearrowright}=&-\left(i\Delta_{c}+\kappa_{0}+\kappa_{ex}\right)\hat{a}_{\circlearrowright}-iJ\hat{a}_{\circlearrowleft}+iG_{0}\hat{a}_{\circlearrowright}\hat{q} \\ \nonumber
 &+\varepsilon_{\circlearrowright}e^{-i\theta_{\circlearrowright}}+\sqrt{2\kappa_{0}}\hat{a}_{0,\circlearrowright}^{\textrm{in}}+\sqrt{2\kappa_{ex}}\hat{a}_{ex,\circlearrowright}^{\textrm{in}},\\ \nonumber
\dot{\hat{a}}_{\circlearrowleft}=&-\left(i\Delta_{c}+\kappa_{0}+\kappa_{ex}\right)\hat{a}_{\circlearrowleft}-iJ\hat{a}_{\circlearrowright}+iG_{0}\hat{a}_{\circlearrowleft}\hat{q}\\ \nonumber
&+\varepsilon_{\circlearrowleft}e^{-i\theta_{\circlearrowleft}}+\sqrt{2\kappa_{0}}\hat{a}_{0,\circlearrowleft}^{\textrm{in}}+\sqrt{2\kappa_{ex}}\hat{a}_{ex,\circlearrowleft}^{\textrm{in}},\\ \nonumber
	\dot{\hat{q}}=&~\omega_{m}\hat{p},\\
	\dot{\hat{p}}=&-\omega_{m}\hat{q}+G_{0}(\hat{a}_{\circlearrowright}^{\dagger}\hat{a}_{\circlearrowright}+\hat{a}_{\circlearrowleft}^{\dagger}\hat{a}_{\circlearrowleft})-\gamma_{m}\hat{p}+\hat{\xi},
	\label{eq2}
\end{align}
where $\gamma_{m}$ is the mechanical damping rate, and $\kappa_{0}$ is the intrinsic decay rate of the optical modes. $\hat{a}_{j}^{\textrm{in}}$ is the optical input vacuum noise operator, whereas $\hat{\xi}$ is the mechanical Brownian noise operator, which are zero mean and characterized by the following correlation functions~\cite{Gardiner2000}:
\begin{align}
	\langle \hat{a}_{j}^{\textrm{in}}\left(t\right)\hat{a}_{j}^{\textrm{in},\dagger}(t^{\prime})\rangle &=\delta(t-t^{\prime}),\notag\\
\left\langle\hat{\xi}(t) \hat{\xi}\left(t^{\prime}\right)\right\rangle&=\frac{\gamma_m}{\omega_m} \int \frac{d \omega}{2 \pi} \mathrm{e}^{-\mathrm{i} \omega\left(t-t^{\prime}\right)}{\omega}\left[\operatorname{coth}\left(\frac{\hbar \omega}{2 k_{\mathrm{B}} T}\right)+1\right].
\end{align}
The Brownian noise operator $\hat{\xi}(t)$, describing stochastic Brownian force acting on mechanical mirrors, is not delta-correlated and generally involves a non-Markovian process. However, in the limit of high quality factor of mechanical mode, i.e., $Q_{m}=\omega_{m}/\gamma_{m}\!\gg\!1$, the correlation function of $\hat{\xi}(t)$ could be reduced to a delta-correlated form as
\begin{align}
\langle\hat{\xi}(t)\hat{\xi}(t^{\prime})\rangle\simeq\gamma_{m}(2n_{m}+1)\delta(t-t^{\prime}),
\end{align}
where $ n_{m}=[\exp(\hbar\omega_{m}/k_{\textit{B}}T)-1]^{-1} $ is the mean thermal excitation phonon number, with $ k_{\textit{B}} $ the Boltzmann constant and $T$ the bath temperature of the mechanical mode.

\begin{figure*}
	\centerline{\includegraphics[width=0.96\textwidth]{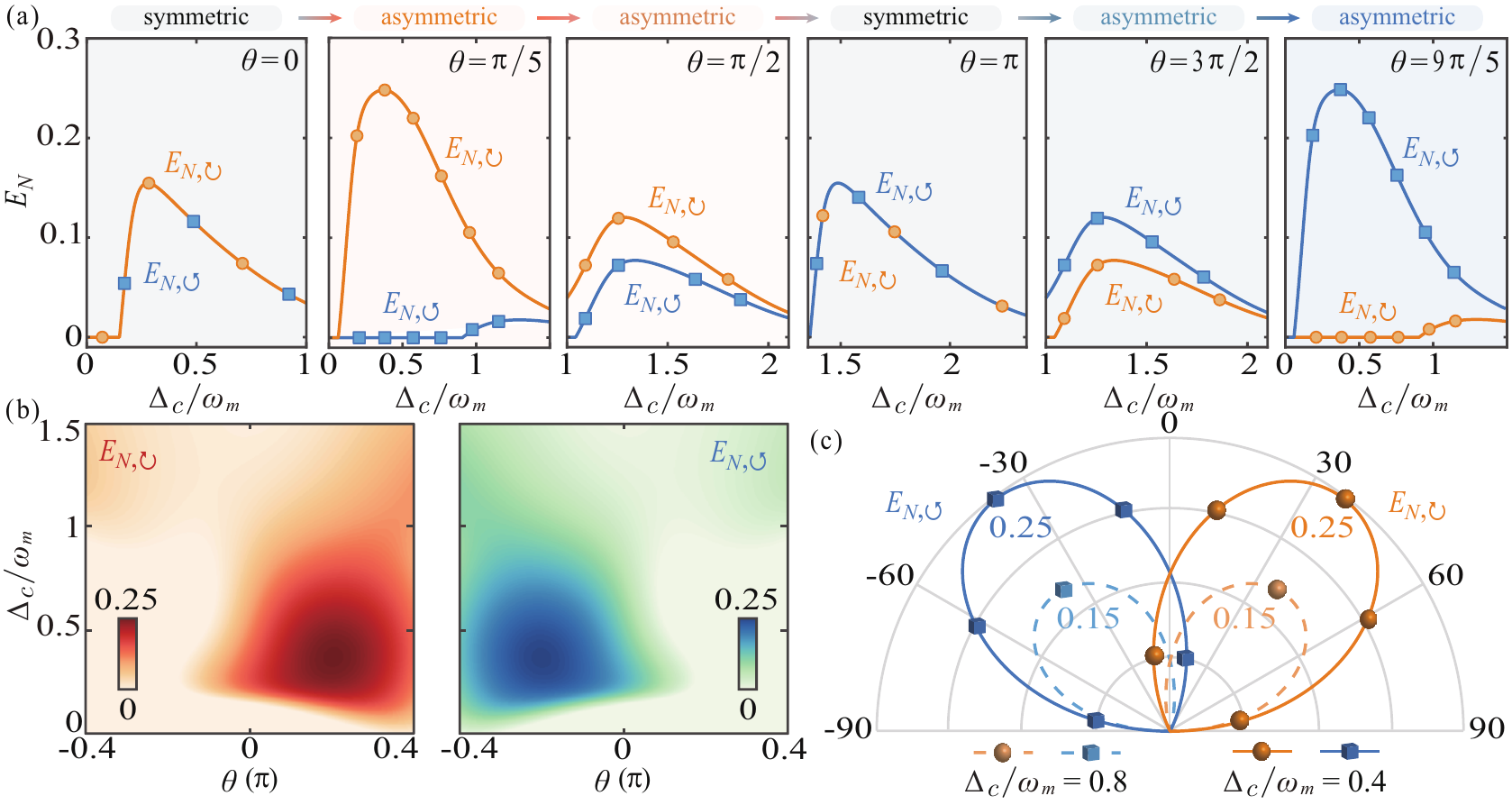}}
	\caption{Coherent asymmetric control of COM entanglement through tuning phase difference of driving fields. (a) The logarithmic negativity $E_{N,j}$ versus the scaled optical detuning $\Delta_{c}/\omega_{m}$ for different values of phase difference $\theta$. The COM entanglement with respect to CW and CCW modes demonstrates a complementary distribution with the variation of $\theta$. (b) Density plot of the logarithmic negativity $E_{N,j}$ as a function of the scaled optical detuning $\Delta_{c}/\omega_{m}$ and the phase difference $\theta$. By adjusting the phase difference $\theta$, the coherent asymmetric control of COM entanglement is achieved. (c) The logarithmic negativity $E_{N,j}$ is plotted as a function of phase difference $\theta$ in polar coordinates, with $\Delta_{c}/\omega_{m}=0.4$ or $0.8$.}\label{Fig3}
\end{figure*}

The dynamics of QLEs involves the nonlinear COM interactions. Under the conditions of strong optical driving, each operator can be expanded as a sum of its steady-state mean value and a small quantum fluctuation around it, i.e.,
\begin{align}
\hat{a}_{j}=\alpha_{j}+\delta\hat{a}_{j},~\hat{q}=q_{s}+\delta\hat{q},~\hat{p}=p_{s}+\delta\hat{p}.
\end{align}
 Therefore, by inserting the above assumptions into Eq.\,(\ref{eq2}), one can linearize the QLEs. Defining the optical quadrature fluctuations
 \begin{align}
\delta \hat{X}_{j}=(\delta \hat{a}_{j}^{\dagger}+\delta \hat{a}_{j})/\sqrt{2},~ \delta \hat{Y}_{j}=i(\delta \hat{a}_{j}^{\dagger}-\delta \hat{a}_{j})/\sqrt{2},
\end{align}
for the optical modes, and
\begin{align}
 \nonumber
&\hat{x}_{j}^{\textrm{in}}=(\hat{a}_{j}^{\textrm{in}\dagger}+\hat{a}_{j}^{\textrm{in}})/\sqrt{2} ,~ \hat{y}_{j}^{\textrm{in}}=i(\hat{a}_{j}^{\textrm{in}\dagger}-\hat{a}_{j}^{\textrm{in}})/\sqrt{2},\\
&\hat{X}_{j}^{\textrm{in}}=\sqrt{2\kappa_{0}}\hat{x}_{0,j}^{\textrm{in}}+\sqrt{2\kappa_{ex}}\hat{x}_{ex,j}^{\textrm{in}},~ \hat{Y}_{j}^{\textrm{in}}=\sqrt{2\kappa_{0}}\hat{y}_{0,j}^{\textrm{in}}+\sqrt{2\kappa_{ex}}\hat{y}_{ex,j}^{\textrm{in}},
\end{align}
for the input vacuum noise, the linearized QLEs could be written explicitly in a compact form, i.e.,
\begin{align}
	\dot{u}(t)=Au(t)+v(t),\label{eq5}
\end{align}
where $u(t)=(\delta \hat{X}_{\circlearrowright},\,\delta \hat{Y}_{\circlearrowright},\,\delta \hat{X}_{\circlearrowleft},\,\delta \hat{Y}_{\circlearrowleft},\,\delta \hat{q},\,\delta \hat{p})^{\text{T}}$, $v(t)=(\hat{X}_{\circlearrowright}^{\textrm{in}},\hat{Y}_{\circlearrowright}^{\textrm{in}},\hat{X}_{\circlearrowleft}^{\textrm{in}},\hat{Y}_{\circlearrowleft}^{\textrm{in}},0,\hat{\xi})^{\text{T}} $, the coefficient matrix $ A $ is given by
\begin{align}
A=\begin{pmatrix}
-\Gamma & \Delta & 0 & J & -G_{\circlearrowright}^{y} & 0\\
-\Delta & -\Gamma & -J & 0 & G_{\circlearrowright}^{x} & 0\\
0 & J & -\Gamma & \Delta & -G_{\circlearrowleft}^{y} & 0\\
-J & 0 & -\Delta & -\Gamma & G_{\circlearrowleft}^{x} & 0\\
0 & 0 & 0 & 0 & 0 & \omega_{m}\\
G_{\circlearrowright}^{x} & G_{\circlearrowright}^{y} & G_{\circlearrowleft}^{x} & G_{\circlearrowleft}^{y} & -\omega_{m} & -\gamma_{m}
\end{pmatrix},
\label{eq6}
\end{align}
where $ \Delta=\Delta_{c}-G_{0}q_{s} $ is the effective optical detuning, and $\Gamma=\kappa_{0}+\kappa_{ex}$ is the total decay rate of the optical modes.
Also, $ G_{j}^{x} $ ($ G_{j}^{y} $) is the real (imaginary) part of the effective COM coupling rate, i.e., $G_{j}\equiv\sqrt{2}G_{0}\alpha_{j}=G_{j}^{x}+iG_{j}^{y}$, where the steady-state mean values of the optical and mechanical modes are given by

\begin{align}
 \nonumber \alpha_{\circlearrowright}&=\frac{e^{-i\theta_{\circlearrowleft}}\left [ \varepsilon_{\circlearrowright}e^{-i\theta}\left(i\Delta+\Gamma\right)-iJ\varepsilon_{\circlearrowleft} \right ]}{\left(i\Delta+\Gamma\right)^{2}+J^{2}},\\
\nonumber
\alpha_{\circlearrowleft}&=\frac{e^{-i\theta_{\circlearrowleft}}\left [\varepsilon_{\circlearrowleft}\left(i\Delta+\Gamma\right)-iJ\varepsilon_{\circlearrowright}e^{-i\theta}\right ]}{\left(i\Delta+\Gamma\right)^{2}+J^{2}},\\
\nonumber
q_{s}&=\dfrac{G_{0}\left(|\alpha_{\circlearrowright}|^{2}+|\alpha_{\circlearrowleft}|^{2}\right)}{\omega_{m}},\\
p_{s}&=0.
\end{align}
Clearly, the intracavity photon numbers, $\mathcal{N}_{j}\equiv|\alpha_{j}|^{2}$, are dominated by the phase difference $\theta$ instead of the phase $\theta_{j}$ of each driving laser.

\begin{figure}
	\centering
	\includegraphics[scale=1]{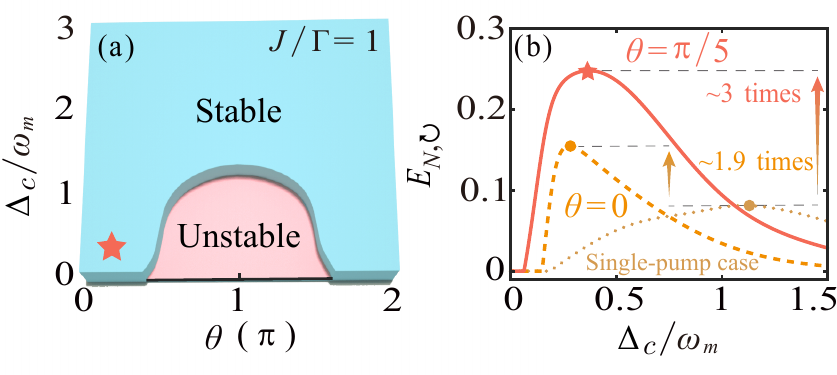}
	\caption{(a) Stability analysis of the system parameters. Density plot of the polynomial equation $\Lambda_{6}$ as a function of the phase difference $\theta$ and the scaled optical detuning $\Delta_{c}/\omega_{m}$. (b) The logarithmic negativity $E_{N,\circlearrowright}$ versus the scaled optical detuning $\Delta_{c}/\omega_{m}$, with $J/\Gamma=1$, and $\theta=0$ or $\pi/5$. In the presence of optical backscattering, compared with the single-pump case, the maximum value of $E_{N,\circlearrowright}$ could be enhanced for $\sim1.9$ or $3$ times for the double-pump case.} 
	\label{Fig4}
\end{figure}

Moreover, the solution of the linearized QLEs could be obtained as
\begin{align}
	u(t)=\mathcal{M}(t)u(0)+\int_{0}^{t}d\tau \mathcal{M}(\tau)v(t-\tau),
\end{align}
where
\begin{align}
	\mathcal{M}(t)=\exp(At).
\end{align}
The system is stable only when all real part of the eigenvalues of $A$
is negative, which can be checked by employing the Routh-Hurwitz criterion~\cite{dejesus1987RouthHurwitza} [see more details in Fig.\,\ref{Fig4}(a)]. When all the stability conditions are fulfilled, one can obtain $\mathcal{M}(\infty)=0$ in the steady state, and
\begin{align}
u_{i}(\infty)=\int_{0}^{\infty}d\tau\sum_{k}\mathcal{M}_{ik}(\tau)v_{k}(t-\tau).\label{eq7}
\end{align}
Because of the linearized dynamics and the Gaussian nature of the input quantum noise, the steady state of the quantum fluctuations of this system finally can evolve into a zero-mean tripartite Gaussian state, which is characterized by a $6\times6$ correlation matrix (CM) $V$, with its matrix components defined as
\begin{align}
	V_{kl}=\langle u_{k}(\infty)u_{l}(\infty)\!+\!u_{l}(\infty)u_{k}(\infty)\rangle/2. \label{eq8}
\end{align}

By substituting Eq.\,(\ref{eq7}) into Eq.\,(\ref{eq8}) and using the fact that the components of $\hat{v}(t)$ are not correlated with each other, the steady-state CM $ V $ is given by
\begin{align}
	V=\int_{0}^{\infty}d\tau\mathcal{M}(\tau)D\mathcal{M}^{\text{T}}(\tau), \label{eq9}
\end{align}
where
\begin{align}
    D\!=\!\textrm{Diag}\,[\Gamma,\Gamma,\Gamma,\Gamma,0,\gamma_{m}(2n_{m}\!+\!1)], \label{eq10}
\end{align}
is obtained by using $ \langle v_{k}(\tau)v_{l}(\tau')\!+\!v_{l}(\tau')v_{k}(\tau)\rangle/2=D_{kl}\delta(\tau-\tau') $. When the stability condition is satisfied, the steady-state CM $V$ is determined by the following Lyapunov equation~\cite{vitali2007Optomechanical}:
\begin{align}
	AV+VA^{\text{T}}=-D. \label{eq11}
\end{align}
Equation (\ref{eq11}) is a linear equation and allows us to derive CM $V$ for any relevant parameter. However, the analytical expression of CM $V$ is too complicated and thus would not be reported here.

\section{Results and discussions}\label{sec:result}

For quantifying bipartite entanglement in a three-mode continuous variable system, one can adopt the logarithmic negativity, $E_{N}$, as an effective measure, which is defined as~\cite{Adesso2004},
\begin{align}
	E_{N}=\max\,[0,-\ln (2\nu^{-})],\label{eq12}
\end{align}
where
\begin{align}
	\nu^{-}\!=\!2^{-1/2}\{\Sigma(V^{\prime})-[\Sigma(V^{\prime})^{2}-4\det\!V^{\prime}]^{1/2}\}^{1/2},
\end{align}
with $ \Sigma(V^{\prime})\!=\!\det\mathcal{A}+\det\mathcal{B}-2\det\mathcal{C} $, is the smallest symplectic eigenvalue of the partial transpose of a reduced $4\times4$ CM $ V^{\prime} $. The reduced CM $ V^{\prime} $ contains the entries of $V$, and it can be directly derived by selecting the rows and columns of the interesting bipartition. By writing the reduced CM $V^{\prime}$ in a $ 2\times2 $ block form, we have
\begin{align}
	V^{\prime}=\left(
	\begin{matrix}
		\mathcal{A}&\mathcal{C}\\
		\mathcal{C}^{\text{T}}&\mathcal{B}
	\end{matrix}
	\right).\label{eq13}
\end{align}
Equation\,(\ref{eq12}) implies that COM entanglement would emerge when $ \nu^{-}<1/2 $, which is equivalent to Simon's necessary and sufficient entanglement nonpositive partial transpose criterion (or the related Peres-Horodecki criterion) for certifying bipartite entanglement in Gaussian states~\cite{simon2000PeresHorodecki}.

\begin{figure}
	\centerline{\includegraphics[width=0.48\textwidth]{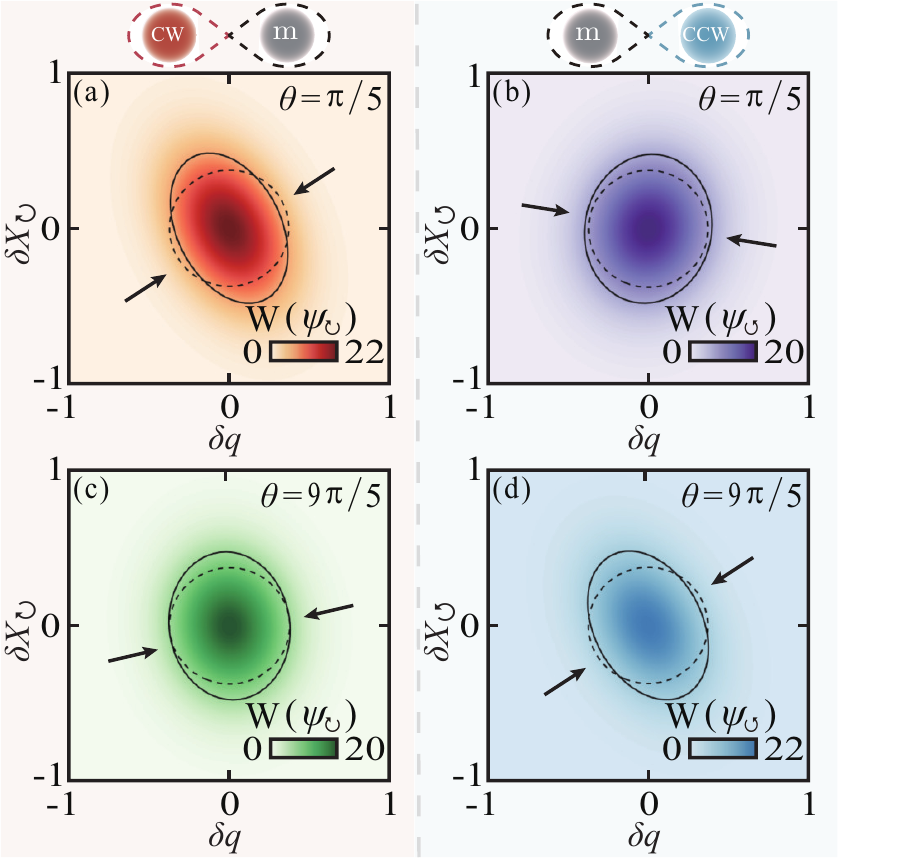}}
	\caption{Coherent switch of two-mode squeezing for cross-quadrature pairs through tuning phase difference $\theta$. The projection of reconstructed Wigner function is plotted on two-dimensional subspaces for cross-quadrature pairs (a), (c) $ \{\delta q, \delta X_{\circlearrowright}\} $ and (b), (d) $ \{\delta q, \delta X_{\circlearrowleft}\} $, with $\Delta_{c}/\omega_{m}=0.4$, and $\theta=\pi/5$ or $9\pi/5$. The ellipse (circle) with solid (dashed) line indicates a drop by $1/\textrm{e}$ of the maximum value of $ \textrm{W}(\psi_{j}) $ for the relevant steady (vacuum) state of the corresponding subsystem.}\label{Fig5}
\end{figure}

In Figs.\,\ref{Fig2} and \,\ref{Fig3}, we first demonstrate how to achieve a coherent asymmetric control of COM entanglement by using synthetic gauge field. As shown in Refs.\,\cite{chen2021Synthetica,Sun2017}, when two optical modes and a mechanical mode are designed to interact with each other to constitute a three-mode closed loop COM system, it would provide an interfering channel to regulate the COM interactions by tuning the relative phase of the driving lasers. By using such configurations, phase control of optical effects or quantum coherence, such as synthetic gauge field~\cite{chen2021Synthetica}, optical nonreciprocity~\cite{Xu2015}, and photon blockade~\cite{Xu2020}, has been intensively investigated in recent studies. In particular, we note that as shown in Ref.\,\cite{Sun2017}, phase control of quantum steering has been explored in blue detuned regime. Here, in comparison with this work~\cite{Sun2017}, we mainly focus on the asymmetric manipulation of the steady-state COM entanglement in red detuned regime and also its counterintuitive property of being robust against optical backscattering.

In our numerical calculation, for ensuring the stability and experimental feasibility of our system, the following parameters are employed: $\omega_{m}=63~\mathrm{MHz}$, $\gamma_{m}=500~\mathrm{Hz}$, $T=130~\mathrm{mK}$, $m=10~\mathrm{ng}$, $\lambda =1550~\mathrm{nm}$, $Q_{c}=\omega_{c}/\kappa_{0}=6.4\times 10^7$, $P=28~\mathrm{mW}$, and $R=1.1~\mathrm{mm}$. Specifically, as shown in Fig.\,\ref{Fig2}, the logarithmic negativity $E_{N,j}$ is plotted as a function of the scaled optical detuning $\Delta_{c}/\omega_{m}$ for different values of optical coupling strength $J$ in regards to single- or double-pump laser case. In terms of applying a single-pump laser, e.g., $\varepsilon_{\circlearrowright}\neq0$ and $\varepsilon_{\circlearrowleft}=0$, $E_{N,\circlearrowright}$ and $E_{N,\circlearrowleft}$ demonstrate a complementary distribution with the variation of optical coupling strength $J$. Particularly, it is shown that the COM entanglement with respect to the driven optical mode tends to be inhibited as $J$ increases [see Fig.\,\ref{Fig2}(a)], implying that optical backscattering is harmful for quantum systems to preserve its coherence. In contrast, for the case of double-pump lasers, i.e., $\varepsilon_{\circlearrowright}=\varepsilon_{\circlearrowleft}\neq0$, both $E_{N,\circlearrowright}$ and $E_{N,\circlearrowleft}$ are considerably enhanced, reaching almost that as in ideal COM resonator with a single-pump laser [see Figs.\,\ref{Fig2}(c) and \ref{Fig2}(d)], which is reminiscent of that in an open loop COM system driven by two-tone driving lasers~\cite{Wang2013abc}.

\begin{figure}
	\centerline{\includegraphics[width=0.48\textwidth]{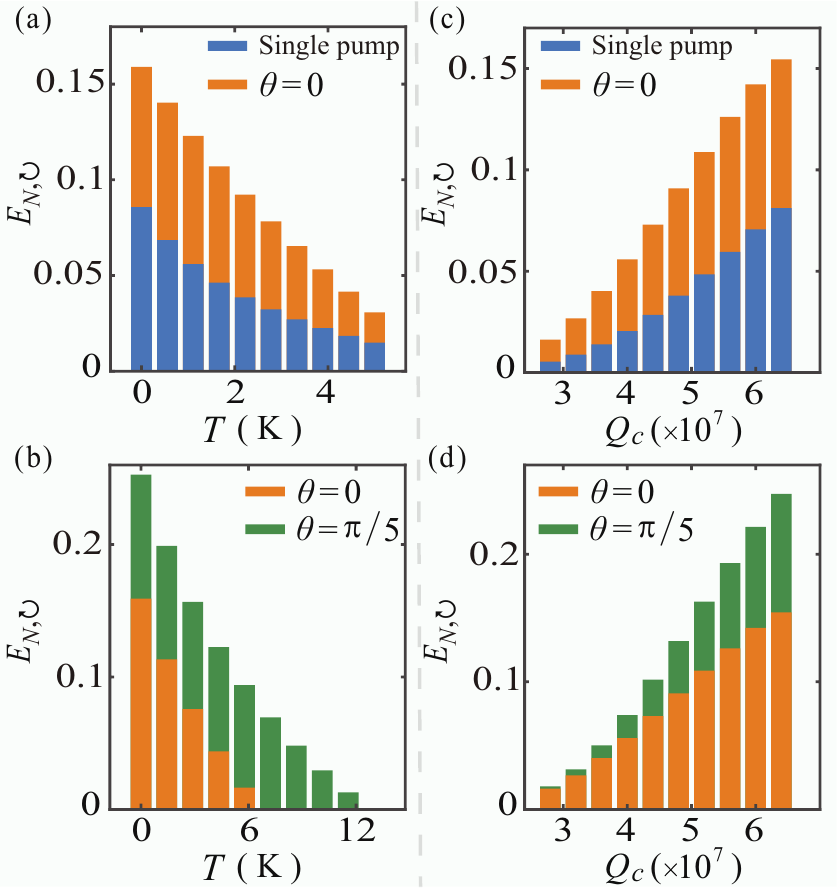}}
	\caption{The influence of thermal effects and optical quality factor $Q_{c}$ on entanglement. (a), (b) The logarithmic negativity $E_{N,\circlearrowright}$ is plotted as a function of the bath temperature $T$. (c), (d) The logarithmic negativity $E_{N,\circlearrowright}$ is plotted as a function of the optical quality factor $Q_{c}$. Here we have chosen $\Delta_{c}/\omega_{m}=1.1$ for single-pump case, $\Delta_{c}/\omega_{m}=0.27$ for double-pump case with $\theta=0$, and $\Delta_{c}/\omega_{m}=0.4$ for double-pump case with $\theta=\pi/5$, which enables the optical value of $E_{N,\circlearrowright}$.}\label{Fig6}
\end{figure}

Furthermore, as shown in Fig.\,\ref{Fig3}, the logarithmic negativity $E_{N,j}$ is plotted as a function of the scaled optical detuning $\Delta_{c}/\omega_{m}$ for different values of phase difference $\theta$ in regards to double-pump laser case. As demonstrated in a very recent experiment~\cite{chen2021Synthetica}, with the same configuration in this paper, a controllable magnetic flux of synthetic gauge field was obtained by tuning the phase difference $\theta$ of two pump lasers, and, simultaneously, a nonreciprocal conversion was also observed through varying this magnetic flux. Figures\,\ref{Fig3}(a)-\ref{Fig3}(c) show the dependence of COM entanglement on phase difference. It is found that in the presence of optical backscattering with $J/\Gamma=1$, $E_{N,\circlearrowright}$ and $E_{N,\circlearrowleft}$ also demonstrate a complementary distribution with the variation of $\theta$, which implies that a coherent asymmetric switch of COM entanglement could be implemented by properly regulating the phase difference. More importantly, as shown in Fig.\,\ref{Fig4}, for some specific values of $\theta$, e.g., $\theta=0$ or $\pi/5$, it is also found that the maximum value of $E_{N,\circlearrowright}$ could be enhanced for $\sim1.9$ or $3$ times in comparison with that of a single-pump driven COM resonator with optical backscattering. The underlying physics can be understood as follows: The WGM resonator could support two counter-propagating optical modes and a mechanical breathing mode. These optical and mechanical modes are coupled with each other via backscattering and optical-radiation-pressure mediated interactions, which results in an effective three-mode closed loop system. Due to these cyclic couplings, optical interference can be created in this closed loop system, when applying two driving fields to it. Thus the field populations and the effective optomechanical couplings turn out to be dependent on the relative phase difference of the two driving fields. As already shown in a very recent work~\cite{chen2021Synthetica}, this relative phase difference is formally equivalent to having a synthetic magnetic flux threading the plaquette formed by the optical and mechanical modes. Hence it provides a new strategy to improve the performance of quantum devices by harnessing the power of synthetic gauge field.

Having explored the abilities of synthetic gauge field in the manipulations of COM entanglement, we now investigate how to modulate and switch the associated two-mode squeezing for cross-quadrature pairs. For this purpose, we introduce the quasiprobability Wigner function $ \textrm{W}(\psi_{j}) $ to visualize the degree of the two-mode squeezing, which is defined as~\cite{barzanjeh2019Stationarya,Adesso2004}
\begin{align}
	\left.\textrm{W}(\psi_{j})=\dfrac{\exp\left\{-\frac{1}{2}\left[\psi_{j}\left ( V^{\prime} \right )^{-1}\psi_{j}^{\dagger}\right] \right\}}{\pi^{2}\sqrt{\det V^{\prime}}}, \right. \label{eq14}
\end{align}
where
\begin{align}
	\psi_{j}=(\delta q, \delta p, \delta X_{j}, \delta Y_{j}), \label{eq21}
\end{align}
is the state vector of orthogonal fluctuations.

As shown in Fig.\,\ref{Fig5}, we plot a selection of characteristic projection of the reconstructed Wigner function $ \textrm{W}(\psi_{j}) $. Here, as a specific example, we have taken $\Delta_{c}/\omega_{m}=0.4$, and $\theta=\pi/5$ or $9\pi/5$, while the other parameters are chosen as the same in Fig.\,\ref{Fig3}. Figures \ref{Fig5}(a) and \ref{Fig5}(b) show the characteristic projections of $ \textrm{W}(\psi_{j}) $ for cross-quadrature pairs $ \{\delta q, \delta X_{\circlearrowright}\} $ and $ \{\delta q, \delta X_{\circlearrowleft}\} $, respectively, along with that of an ideal vacuum state for reference. The ellipse (circle) with solid (dashed) line indicates a drop by $1/\textrm{e}$ of the maximum value of $ \textrm{W}(\psi_{j}) $ for the relevant steady (vacuum) state of the corresponding subsystem. It is seen that, in terms of $\theta=\pi/5$, the boundary of the ellipse could enter into the circle for $\{\delta q, \delta X_{\circlearrowright}\}$, while it is otherwise unattainable for $\{\delta q, \delta X_{\circlearrowleft}\}$ under the same conditions, indicating that the two-mode squeezing could only be achieved between the CW and mechanical modes. In contrast, in terms of $\theta=9\pi/5$, the two-mode squeezing is generated between the CCW and mechanical modes. These results mean that by tuning the phase difference of the driving lasers, one can implement coherent switch of two-mode squeezing between the optical and mechanical modes.

Finally, we also confirm the dependence of the logarithmic negativity $E_{N,\circlearrowright}$ on temperature and cavity quality factor with respect to the single- or double-pump case in Fig.\,\ref{Fig6}. As shown in Figs.\,\ref{Fig6}(a) and \ref{Fig6}(b), for both of the single- and double-pump case, the COM entanglement decreases monotonically when the bath temperature increases. Similarly, as shown in Figs.\,\ref{Fig6}(c)-\ref{Fig6}(d), the COM entanglement also tends to be inhibited by the reduction of cavity quality factor. However, it is also seen that, under the same conditions of temperature or cavity quality factor, the degree of COM entanglement could be considerably enhanced for the double-pump case in comparison with that of a single driving laser.

\section{Conclusion}\label{sec:conclusion}
In summary, by using the unique properties of synthetic gauge fields in a single COM resonator, we have shown how to achieve a coherent asymmetric control of COM entanglement in regards to two counter-propagating optical modes, how to implement a switch of the associated two mode squeezing between different cross-quadrature pairs, and how to preserve and even enhance the robustness of quantum coherence against optical backscattering. Our findings, shedding a new light on improving the performance of various quantum devices in practical noisy environment, provide an enticing new opportunity to realize such a wide range of entanglement-enabled quantum technologies as quantum sensing~\cite{Degen2017,Gilmore2021,Barzanjeh2020}, quantum computing~\cite{Gyongyosi2019,Knill2005,OBrien2007}, and quantum networking~\cite{Hermans2022,Simon2017,Komar2014}. In a broader view, the abilities to coherently manipulate light-motion interactions through controlling the magnetic fluxes of a synthetic gauge field could also open up a promising way to engineer various other nonlinear COM effects, such as phonon lasers~\cite{Chafatinos2020,Pettit2019,Jing2014,Vahala2009}, slow light control~\cite{Vlasov2005,DiFalco2008,Baba2008,Wu2010}, or quantum sensing~\cite{Degen2017,Gilmore2021,Barzanjeh2020}. In addition, we emphasize that although we have considered here a specific case of the synthetic gauge field which is realized by tuning the phase difference of driving lasers within COM resonators, future developments with the manipulations of quantum effect via synthetic gauge field could further be extended to various other physical platforms, including atomic gases~\cite{Li2022}, photonic devices~\cite{Umucalilar2011,Lumer2019,Cohen2020}, acoustic or optomechanical systems~\cite{Yang2017ac,Longhi2015}, and topological structures~\cite{Huang2022}.

\section{Acknowledgments}\label{V}
H. J. is supported by the National Natural Science Foundation of China (Grants No. 11935006 and No. 11774086) and the Science and Technology Innovation Program of Hunan Province (Grant No. 2020RC4047). Y.-F. J. is supported by the NSFC (Grant No. 12147156), the China Postdoctoral Science Foundation (Grant No. 2021M701176 and No. 2022T150208) and the Science and Technology Innovation Program of Hunan Province (Grant No. 2021RC2078). Q.-Y. H. is supported by the NSFC (Grants No. 12125402 and No. 11975026). X.-W. X. is supported by the NSFC (Grants No. 12064010), and Natural Science Foundation of Hunan Province of China (Grant No. 2021JJ20036). We appreciate valuable discussions with Chun-Hua Dong.

\begin{appendix}

\section{Stability analysis} \label{appendix A}
In order to study COM entanglement, we need to concern about the stability of the COM system. Here, we can get the stability of the COM system by the eigenvalues of the coefficient matrix A. Then, we can obtain the conditions for the COM system stability by the Routh-Hurwitz criterion~\cite{dejesus1987RouthHurwitza}.  The eigenequation of Eq.\,(\ref{eq6}) is as follows:

\begin{equation}
	A\mathbf{x}=\eta\mathbf{x}, \label{eqA1}
\end{equation}
we can obtain the eigenvalues by solving this equation,
\begin{equation}
	|A-\eta\textbf{I}|=0, \label{eqA2}
\end{equation}
then, the following equation can be derived,
\begin{equation}
	F(\eta)=\sum_{n=0}^{6}a_{6-n}\eta^{n}=0, \label{eqA3}
\end{equation}
with,
\begin{align}
	a_{0}\eta^{6}+a_{1}\eta^{5}+a_{2}\eta^{4}+a_{3}\eta^{3}
	+a_{4}\eta^{2}+a_{5}\eta+a_{6}=0. \label{eqA4}
\end{align}
The corresponding coefficients:
\begin{widetext}
\begin{align}
	\nonumber
a_{0}=&~1,\\ \nonumber
a_{1}=&~4\Gamma+\gamma_{m},\\ \nonumber
a_{2}=&~2J^{2}+2\Delta^{2}+4\gamma_{m}\Gamma+6\Gamma^{2}+\omega_{m}^{2},\\ \nonumber
a_{3}=&~2J^{2}\gamma_{m}+2\gamma_{m}\Delta^{2}+4J^{2}\Gamma+4\Delta^{2}\Gamma+6\gamma_{m}\Gamma^{2}+4\Gamma^{3}+4\Gamma\omega_{m}^{2},\\ \nonumber	
a_{4}=&~J^{4}-2J^{2}\Delta^{2}+\Delta^{4}+4J^{2}\gamma_{m}\Gamma+4\gamma_{m}\Delta^{2}\Gamma+2J^{2}\Gamma^{2}+2\Delta^{2}\Gamma^{2}+4\gamma_{m}\Gamma^{3}-2G_{\circlearrowright}^{x}G_{\circlearrowleft}^{x}J\omega_{m}-2G_{\circlearrowright}^{y}G_{\circlearrowleft}^{y}J\omega_{m} \\ \nonumber
&-\left(G_{\circlearrowleft}^{x}\right)^{2}\Delta\omega_{m}-\left(G_{\circlearrowright}^{x}\right)^{2}\Delta\omega_{m}-\left(G_{\circlearrowleft}^{y}\right)^{2}\Delta\omega_{m}-\left(G_{\circlearrowright}^{y}\right)^{2}\Delta\omega_{m}+\Gamma^{4}+6\Gamma^{2}\omega_{m}^{2}+2\Delta^{2}\omega_{m}^{2}+2J^{2}\omega_{m}^{2},\\ \nonumber
a_{5}=&~J^{4}\gamma_{m}-2J^{2}\gamma_{m}\Delta^{2}+\gamma_{m}\Delta^{4}+2J^{2}\gamma_{m}\Gamma^{2}-4G_{\circlearrowright}^{y}G_{\circlearrowleft}^{y}J\Gamma\omega_{m}+2\gamma_{m}\Delta^{2}\Gamma^{2}+\gamma_{m}\Gamma^{4}-4G_{\circlearrowright}^{x}G_{\circlearrowleft}^{x}J\Gamma\omega_{m}+4J^{2}\Gamma\omega_{m}^{2}\\ \nonumber	
&-2\left(G_{\circlearrowleft}^{x}\right)^{2}\Delta\Gamma\omega_{m}-2\left(G_{\circlearrowright}^{x}\right)^{2}\Delta\Gamma\omega_{m}+4\Delta^{2}\Gamma\omega_{m}^{2}-2\left(G_{\circlearrowleft}^{y}\right)^{2}\Delta\Gamma\omega_{m}-2\left(G_{\circlearrowright}^{y}\right)^{2}\Delta\Gamma\omega_{m}+4\Gamma^{3}\omega_{m}^{2},\\ \nonumber
a_{6}=&~-2G_{\circlearrowright}^{y}G_{\circlearrowleft}^{y}J\Gamma^{2}\omega_{m}-\left(G_{\circlearrowleft}^{x}\right)^{2}\Gamma^{2}\Delta\omega_{m}-2J^{2}\Delta^{2}\omega_{m}^{2}-\left(G_{\circlearrowright}^{y}\right)^{2}\Delta^{3}\omega_{m}-2G_{\circlearrowright}^{x}G_{\circlearrowleft}^{x}J\Gamma^{2}\omega_{m}+\Delta^{4}\omega_{m}^{2}+2J^{2}\omega_{m}^{2}\Gamma^{2}\\ \nonumber
&+2G_{\circlearrowright}^{x}G_{\circlearrowleft}^{x}J\Delta^{2}\omega_{m}+2G_{\circlearrowright}^{y}G_{\circlearrowleft}^{y}J\Delta^{2}\omega_{m}-\left(G_{\circlearrowleft}^{y}\right)^{2}\Delta^{3}\omega_{m}-2G_{\circlearrowright}^{x}G_{\circlearrowleft}^{x}J^{3}\omega_{m}-2G_{\circlearrowright}^{y}G_{\circlearrowleft}^{y}J^{3}\omega_{m}+\left(G_{\circlearrowright}^{y}\right)^{2}J^{2}\Delta\omega_{m}\\ \nonumber
&+\left(G_{\circlearrowleft}^{x}\right)^{2}J^{2}\Delta\omega_{m}+\left(G_{\circlearrowright}^{x}\right)^{2}J^{2}\Delta\omega_{m}+\left(G_{\circlearrowleft}^{y}\right)^{2}J^{2}\Delta\omega_{m}+2\Delta^{2}\Gamma^{2}\omega_{m}^{2}-\left(G_{\circlearrowleft}^{x}\right)^{2}\Delta^{3}\omega_{m}-\left(G_{\circlearrowright}^{x}\right)^{2}\Delta^{3}\omega_{m}+\omega_{m}^{2}\Gamma^{4}\\
&+J^{4}\omega_{m}^{2}-\left(G_{\circlearrowright}^{x}\right)^{2}\Gamma^{2}\Delta\omega_{m}-\left(G_{\circlearrowleft}^{y}\right)^{2}\Gamma^{2}\Delta\omega_{m}-\left(G_{\circlearrowright}^{y}\right)^{2}\Gamma^{2}\Delta\omega_{m}.
\end{align}
\end{widetext}
According to the Routh-Hurwitz criteria, ensuring that all roots of the equation $F(\eta)=0$ are negative or have negative real parts. The determinant that determines the stability of the COM system is as follows:
\begin{widetext}
\begin{align}
		\Lambda_{n}=\left\vert
		\begin{array}{cccccccc}
			a_{1} & 1 & 0 & 0 & 0 & 0 & \ldots & 0 \\
			a_{3} & a_{2} & a_{1} & 1 & 0 & 0 & \ldots & 0 \\
			a_{5} & a_{4} & a_{3} & a_{2} & a_{1} & 1 & \ldots & 0 \\
			\ldots & \ldots & \ldots & \ldots & \ldots & \ldots & \ldots & \ldots \\
			a_{2n-1} & a_{2n-2} & a_{2n-3} & a_{2n-4} & a_{2n-5} & a_{2n-6} & \ldots &
			a_{n}
		\end{array}
		\right\vert>0, \label{eqA6}
\end{align}
\end{widetext}
for $n=1$-$6$. To ensure that the parameters we use in our numerical calculations are in the stability region, in Fig.\,\ref{Fig4}(a) we numerically plot the functions $\Lambda_{6}$ to present the boundary between the stability and the instability regions. Here we only give the result of the smallest value of the determinant in the determinants $\Lambda_{1}$-$\Lambda_{6}$. The minimum value $\Lambda_{6}$ can divid the stability region of the system. In our numerical simulations, we have confirmed that  the parameters chosen in this paper can keep the COM system in a stability regions.

\end{appendix}

\end{document}